\def\rfr#1{eq. (\ref{#1})}

\def\dert#1#2{\frac{{{d}}{#1}}{{{d}}{#2}}}

\def\eqi{\begin{equation}}
\def\eqf{\end{equation}}
\def\eqia{\begin{eqnarray}}
\def\eqfa{\end{eqnarray}}
\def\Om{\mathit{\Omega}}
\def\rp#1#2{{#1\over#2}} \def\lb#1{\label{#1}}

\def\bds#1{\boldsymbol{#1}}

\def\co{\cos\omega}
\def\so{\sin\omega}

\def\cO{\cos\Om}
\def\sO{\sin\Om}

\def\cI{\cos I}
\def\sI{\sin I}

\def\ee{e^2}

\def\ton#1{\left(#1\right)}
\def\qua#1{\left[#1\right]}
\def\grf#1{\left\{#1\right\}}
\def\ang#1{\left\langle #1\right\rangle}

\documentclass[11pt]{article}
\usepackage{url}
\usepackage{amsmath,amsthm,amscd,amssymb}
\usepackage{latexsym,w-greek,wasysym}
\usepackage{graphicx,epsfig}
\usepackage{hyperref}
\allowdisplaybreaks[1]

\RequirePackage{color}

\begin{document}

\title{Orbital effects of Lorentz-violating Standard Model Extension gravitomagnetism around a static body: a sensitivity analysis}

\author{L. Iorio \\ Ministero dell'Istruzione, dell'Universit\`{a} e della Ricerca (M.I.U.R.)-Istruzione \\ Fellow of the Royal Astronomical Society (F.R.A.S.) \\
 International Institute for Theoretical Physics and
Advanced Mathematics Einstein-Galilei \\ Permanent address: Viale Unit$\grave{\rm a}$ di Italia 68
70125 Bari (BA), Italy \\ email: lorenzo.iorio@libero.it}

\maketitle

\begin{abstract}
We analytically work out the long-term rates of change of the six osculating Keplerian orbital elements of a test particle acted upon by the Lorentz-violating gravitomagnetic acceleration due to a static body, as predicted by  the Standard Model Extension (SME). We neither restrict to any specific spatial orientation for the symmetry-violating vector $\bds s=\{-\overline{s}^{01},-\overline{s}^{02},-\overline{s}^{03}\}$ nor make a priori simplifying assumptions concerning the orbital configuration of the perturbed test particle. Thus, our results are quite general, and can be applied for sensitivity analyses to a variety of specific astronomical and astrophysical scenarios.
We find that, apart from the semimajor axis $a$, all the other orbital elements undergo non-vanishing secular variations. By comparing our results to the latest determinations of the supplementary advances of the perihelia of some planets of the solar system we preliminarily obtain $s_x   = (0.9\pm 1.5)\times 10^{-8},
s_y = (-4 \pm 6)\times 10^{-9}, s_z  = (0.3\pm 1)\times 10^{-9}.$ Bounds from the terrestrial LAGEOS and LAGEOS II satellites are of the order of $s\sim 10^{-3}-10^{-4}$.
\end{abstract}

%
\centerline
\section{Introduction}
There are a variety of theoretical schemes predicting violations of Lorentz symmetry. Attempts to quantize gravity have resulted in theories that
allow for it \cite{Ame05}. Certain string theories envisage the possibility of a spontaneous breaking of it \cite{koste89}.
The possibility of detecting experimental signatures of breakings of Lorentz symmetry \cite{Ame00,Ame05} from an underlying unified theory at the Planck scale has recently raised interest in Lorentz violation; for a recent review of modern tests of Lorentz invariance see, e.g., \cite{Matt05}.

The Standard Model Extension (SME) is a theoretical framework which allows for generic violations of the Lorentz symmetry for both gravity and electromagnetism \cite{koste1,koste2,koste3,koste4}. In general, there are 20 coefficients for Lorentz violation in the gravitational sector; by assuming spontaneous Lorentz-symmetry breaking, the main effects in the weak-field approximation are accounted for by the traceless coefficients $\overline{s}^{\mu\nu}$ \cite{Bail06} containing 9 independent quantities.

According to Bailey \cite{Bail010}, in the weak field and slow motion approximation, a test particle moving with velocity $\bds v$ at distance $r$ from a central, static body of mass $M$ experiences a Lorentz-violating gravitomagnetic acceleration
\eqi \bds A_{\rm GM} =\ton{\rp{\bds v}{c}}\bds\times \bds B_{\rm G}, \lb{accel}\eqf
where
\eqi\bds B_{\rm G}\doteq \rp{2GM}{r^3}\ton{\bds s\bds\times \bds r},\lb{BG}\eqf
with \eqi \bds s\doteq -\overline{s}^{0j},\ j=1,2,3;\eqf $G$ is the Newtonian constant of gravitation, while $c$ is the speed of light in vacuum.
Constraints from Lunar Laser Ranging (LLR) \cite{Dick} are \cite{Batt07}
\begin{align}
\overline{s}^{01} \lb{llr1} & = \ton{-0.8\pm 1.1}\times 10^{-6}, \\ \nonumber \\
\overline{s}^{02} \lb{llr2} & = \ton{-5.2\pm 4.8}\times 10^{-7},
\end{align}
so that
\eqi s_{\rm max}\lesssim 1-2\times 10^{-6}.\lb{smax}\eqf
A general account of the present-day bounds on all the SME Lorentz violating coefficients can be found in \cite{Koste011}.

From simple dimensional considerations, planetary orbital precessions in the field of the Sun, if present, would be as large as
\eqi \left|\dot\Psi\right|\lesssim \rp{A_{\rm G}}{v}\sim \rp{2GMs}{cr^2} = \rp{1.77\times 10^6\ {\rm m^2\ s^{-1}}}{r^2}, \lb{prece}\eqf
where $\Psi$ denotes a generic orbital element, and we used \rfr{smax}.
The present-day level of accuracy in constraining the secular rates of orbital changes of the planets of the solar system are listed in Table \ref{tavolapit} and Table \ref{tavolafie}.
\begin{table*}[ht!]
\caption{Uncertainties in the rates of change of the semimajor axis $a$, the eccentricity $e$, the inclination $I$ to the mean ecliptic at J$2000.0$, the longitude of the ascending node $\Om$, the longitude of perihelion $\varpi\doteq \Om + \omega$, the mean motion $n_{\rm b}\doteq\sqrt{GM a^{-3}}$ and the mean longitude $\lambda\doteq\Om+\omega+\mathcal{M}$ of the planets of the solar system; $\omega$ is the argument of perihelion, while $\mathcal{M}$ is the mean anomaly. They were inferred by us by taking the ratios of the formal, statistical errors in Table 3 of \cite{pitjeva07}, all rescaled by a factor 10, to the data time span $\Delta T=93$ yr (1913-2006) of the EPM2006 ephemerides used by Pitjeva \cite{pitjeva07}. The figures for $n_{\rm b}$ account also for the uncertainty $\sigma_{GM}=10\ {\rm km^3\ s^{-2}}$ in the Sun's gravitational parameter $GM$ retrieved from \cite{konopliv011}. The results for Saturn are relatively inaccurate with respect to those of the inner planets since radiotechnical data from Cassini were not yet processed when Table 3 of \cite{pitjeva07} was produced. Here mas cty$^{-1}$ stands for milliarcseconds per century.
}\label{tavolapit}
\centering
\bigskip
\begin{tabular}{llllllll}
\hline\noalign{\smallskip}
& $\dot a$ $\ton{\rp{\rm m}{\rm cty}}$ & $\dot e$ $\ton{\rp{1}{\rm cty}}$ & $\dot I$ $\ton{\rp{\rm mas}{\rm cty}}$ &  $\dot \Om$ $\ton{\rp{\rm mas}{\rm cty}}$ & $\dot \varpi $ $\ton{\rp{\rm mas}{\rm cty}}$ & $n_{\rm b}$ $\ton{\rp{\rm mas}{\rm cty}}$ & $\dot \lambda$ $\ton{\rp{\rm mas}{\rm cty}}$ \\
\noalign{\smallskip}\hline\noalign{\smallskip}
Mercury & $3.6$ & $4\times 10^{-9}$ & $14.8$ & $121$ & $5.4$ & $50.6$ & $3.7$ \\
Venus & $2.3$ & $2\times 10^{-10}$ & $0.3$ &  $9.9$ & $5.7$ & $10.2$ &  $0.3$\\
Earth & $1.5$ & $5\times 10^{-11}$ & $-$ &  $-$ & $0.6$ & $5.2$ & $-$ \\
Mars & $2.8$ & $5\times 10^{-11}$ & $0.03$ &  $0.8$ & $0.1$ & $2.8$ &  $0.02$\\
Jupiter & $6612.9$ & $2\times 10^{-8}$ & $23.4$ &  $1138.9$ & $79.8$ & $131.2$ &  $15.7$\\
Saturn & $45763.4$ & $1\times 10^{-7}$ & $42.7$ &  $806.7$ & $778.2$ & $196.9$ &  $37.6$\\
Uranus & $433269.0$ & $3\times 10^{-7}$ & $66.2$ &  $3671.9$ & $836.5$ & $323.7$ &  $80.7$\\
Neptune & $4.9818\times 10^6$ & $8\times 10^{-7}$ & $75.3$ &  $2284.0$ & $20249.1$ & $1210.7$ & $263.2$ \\
Pluto & $3.66961\times 10^7 $ & $3\times 10^{-6}$ & $167.4$ &  $305.4$ &  $2713.4$ & $4501.6$ &  $468.3$\\
\noalign{\smallskip}\hline\noalign{\smallskip}
\end{tabular}
\end{table*}
\begin{table*}[ht!]
\caption{Supplementary advances of
perihelia and nodes of some planets of the solar system estimated by \cite{fienga011} with the INPOP10a ephemerides. Data from Messenger and Cassini were included for Mercury and Saturn. The reference $\{x,y\}$ plane is the mean Earth's equator at J$2000.0$.
}\label{tavolafie}
\centering
\bigskip
\begin{tabular}{lll}
\hline\noalign{\smallskip}
&   $\dot \Om$ $\ton{\rp{\rm mas}{\rm cty}}$ & $\dot \varpi $ $\ton{\rp{\rm mas}{\rm cty}}$  \\
\noalign{\smallskip}\hline\noalign{\smallskip}
Mercury & $1.4 \pm 1.8$ & $0.4 \pm 0.6$ \\
Venus & $0.2 \pm 1.5$ & $ 0.2\pm 1.5$ \\
Earth & $0.0\pm 0.9$ & $-0.2\pm 0.9$ \\
Mars & $-0.05\pm 0.13$ & $-0.04\pm 0.15$ \\
Jupiter & $-40\pm 42$ & $-41\pm 42$ \\
Saturn & $-0.1\pm 0.4$ & $0.15\pm 0.65$ \\
\noalign{\smallskip}\hline\noalign{\smallskip}
\end{tabular}
\end{table*}
In Table \ref{tavolaSME} we show tentative upper bounds on putative precessional effects according to \rfr{smax} for all the planets of the solar system.
\begin{table*}[ht!]
\caption{Orders of magnitude of the maximum values $\left|\dot \Psi\right|_{\rm max}$ of putative planetary orbital precessions due to \rfr{accel}, with $s = s_{\rm max}\sim 2\times 10^{-6}$. Cfr. with the empirical bounds in Table \ref{tavolapit} and Table \ref{tavolafie}.
}\label{tavolaSME}
\centering
\bigskip
\begin{tabular}{lll}
\hline\noalign{\smallskip}
&   $\left|\dot \Psi\right|_{\rm max}$ $\ton{\rp{\rm mas}{\rm cty}}$ & $\left|\dot \Psi\right|_{\rm max} $ $\ton{\rp{1}{\rm cty}}$  \\
\noalign{\smallskip}\hline\noalign{\smallskip}
Mercury & $343.70$ & $1.7\times 10^{-6}$ \\
Venus & $98.43$ & $4.7\times 10^{-7}$ \\
Earth & $51.50$ & $2.5\times 10^{-7}$ \\
Mars & $22.18$ & $1.1\times 10^{-7}$ \\
Jupiter & $1.92$ & $9.3\times 10^{-9}$ \\
Saturn & $0.56$ & $2.7\times 10^{-9}$ \\
Uranus & $0.14$ & $6.7\times 10^{-10}$\\
Neptune & $0.05$ & $2.7\times 10^{-10}$ \\
Pluto & $0.03$ & $1.6\times 10^{-10}$ \\
\noalign{\smallskip}\hline\noalign{\smallskip}
\end{tabular}
\end{table*}
A tension between the LLR bounds \cite{Batt07} of  \rfr{llr1}-\rfr{llr2} and the resulting expected planetary precessions of Table \ref{tavolaSME} exist, especially for the inner planets.

Thus, a more detailed analysis is worthwhile. Actually,  a mere order-of-magnitude analysis based just on \rfr{prece} and Table \ref{tavolaSME} would be insufficient to draw meaningful
conclusions. Indeed, exact calculations of the secular variations of all the osculating Keplerian orbital elements caused by \rfr{accel} must
be performed, with, e.g., standard perturbative techniques, in order to check if it really induces  averaged non-zero orbital changes.
Moreover, also in such potentially favorable case caution is still in order. Indeed, it may well happen, in principle, that the resulting
analytical expressions retain multiplicative factors $1/e^k, k = 1, 2, 3,\ldots$ or $e^k, k = 1, 2, 3,\ldots $ which would notably alter the size of the
found non-zero secular rates with respect to the expected values according to \rfr{prece}. Thus, in Section \ref{calcolo}, we will analytically work out the long-term rates of change of all the  osculating Keplerian orbital elements of a test particle acted upon by \rfr{accel} by using the Gauss perturbative scheme \cite{roy05}.
\section{Analytical calculation of the orbital precessions}\lb{calcolo}
In the following, we will not make any particular assumption about the orientation of $\bds s$ in space, so that we will adopt a generic reference frame centered in $M$, with respect to which $\bds s=\{s_x,s_y,s_z\}$. Moreover, we will not make any definite choice about the reference $\{x,y\}$ plane, so that our results can be applied to a variety of specific astronomical and astrophysical scenarios.

The\footnote{The radial unit vector $\bds{\hat{R}}$ is directed from $M$ to the test particle, the transverse unit vector $\bds{\hat{T}}$ lies in the osculating orbital plane and is perpendicular to $\bds{\hat{R}}$, while the normal unit vector $\bds{\hat{N}}$ is perpendicular to the orbital plane and is directed along the osculating orbital angular momentum $\bds L$. Such three unit vectors constitute  a right-handed orthonormal basis comoving with the test particle.} radial, transverse and normal components $A_R,A_T,A_N$ of \rfr{accel}, evaluated onto the unperturbed Keplerian ellipse, are
\begin{align}
A_R \lb{ar} & = \rp{-2 GM n_{\rm b} \ton{1 + e \cos f}^3 }{ a c \ton{1 -e^2}^{5/2} } \mathcal{A_R}, \\ \nonumber \\
\mathcal{A_R} & = s_z \cos u \sin  I  + \cos  I  \cos u \ton{s_y \cos \Om  - s_x \sin \Om } - \sin u \ton{s_x \cos \Om  + s_y \sin \Om }, \\ \nonumber \\
A_T & = \rp{2 e GM n_{\rm b} \ton{1 + e \cos f}^2 \sin f }{ a c \ton{1 - e^2}^{5/2}}\mathcal{A_T}, \\ \nonumber \\
\mathcal{A_T} & = s_z \cos u\sin  I  + \cos  I  \cos u \ton{s_y \cos \Om  - s_x
\sin \Om } - \sin u \ton{s_x \cos \Om  + s_y \sin \Om }, \\ \nonumber \\
A_N \lb{an} & = \rp{2 e GM n_{\rm b} \ton{1 + e \cos f}^2 \sin f }{a c \ton{1 - e^2}^{5/2}} \mathcal{A_N}, \\ \nonumber \\
\mathcal{A_N}  & = s_z \cos  I  + \sin  I\ton{s_x \sin \Om -s_y \cos \Om},
\end{align}
where \cite{roy05}  $u\doteq f + \omega$ is the argument of latitude defined in terms of the true anomaly $f$ and the argument of pericenter $\omega$. Notice that $A_T = A_N = 0$ for circular orbits, i.e. for $e\rightarrow 0$.
Inserting \rfr{ar}-\rfr{an} in the right-hand-sides of the Gauss equations for the variation of the elements and averaging them over one orbital revolution yield the following non-vanishing rates of changes of the Keplerian orbital elements
\begin{align}
\ang{\dert a t} \lb{dadt} & = 0, \\ \nonumber \\
\ang{\dert e t} \lb{dedt} & =\rp{ 2GM\sqrt{1-\ee}}{ c a^2 \ton{ 1 + \sqrt{1-\ee} } }\mathcal{E}, \\ \nonumber \\
\mathcal{E} & = \co \ton{s_x \cO + s_y \sO} + \so \qua{s_z \sI + \cI \ton{s_y \cO - s_x \sO}}, \\ \nonumber \\
\ang{\dert I t} \lb{dIdt} & = -\rp{ 2GM\ton{ 1 - \sqrt{1-\ee}}}{ c a^2 e\sqrt{1-\ee} }\mathcal{I},\\ \nonumber \\
\mathcal{I} & = \so\qua{s_z \cI +\sI\ton{s_x\sO - s_y\cO}}, \\ \nonumber \\
\ang{\dert \Om t} \lb{dOdt} & = \rp{ 2GM\ton{ 1 - \sqrt{1-\ee}}}{ c a^2 e\sqrt{1-\ee} }\mathcal{N},\\ \nonumber \\
\mathcal{N} & = \co\csc I\qua{s_z\cI +\sI\ton{s_x\sO - s_y\cO} }, \\ \nonumber \\
\ang{\dert \varpi t} \lb{dodt} & = -\rp{ 2GM}{ c a^2 e^3\ton{1-e^2} }\mathcal{P},\\ \nonumber \\
\mathcal{P} \nonumber & = \ton{1 - e^2} \ton{1 - \sqrt{1 - e^2}} \sin\omega \ton{s_x \cos\Om + s_y \sin\Om} + \\ \nonumber \\
\nonumber & +  \cos\omega \grf{s_y \qua{e^2 \ton{-1 + e^2 + \sqrt{1 - e^2}} + \right.\right. \\ \nonumber \\
\nonumber & + \left.\left. \ton{-1 + \sqrt{1 - e^2} - e^2 \ton{-2 + e^2 + 2 \sqrt{1 - e^2}}} \cos I} \cos\Om +\right.\\ \nonumber \\
\nonumber & + \left.  \ton{1 - e^2} \ton{1 - \sqrt{1 - e^2}} s_z \sin I - e^2 \sqrt{1 - e^2} s_x \sin\Om - \sqrt{1 - e^2} s_x
\cos I \sin\Om + \right. \\ \nonumber \\
\nonumber & + \left. 2 e^2 \sqrt{1 - e^2} s_x \cos I \sin\Om + \ton{-1 + e^2} s_x \qua{-e^2 + \ton{-1 + e^2} \cos I} \sin\Om -\right.\\ \nonumber \\
& - \left. e^2 \sqrt{1 - e^2} s_z \cos I \tan\ton{\rp{I}{2}} + e^2 \ton{1 - e^2} s_z \cos I \tan\ton{\rp{I}{2}}}, \\ \nonumber \\
\ang{\dert {\mathcal{M}} t} \lb{dMdt} & =-\rp{ 2GM\ton{1 - e^2}}{ c a^2 e\ton{1 - e^2 + \sqrt{1-e^2}} }\mathcal{L},\\ \nonumber \\
\mathcal{L} & = s_z \cos\omega \sin I + \cos I \cos\omega \ton{s_y \cos\Om - s_x \sin\Om} - \sin\omega \ton{s_x \cos\Om + s_y \sin\Om}.
\end{align}
The results of \rfr{dadt}-\rfr{dMdt} are exact in the sense that no a priori simplifying assumptions on either $e$ and $I$ have been assumed.
By expanding in powers of $e$, it turns out that, in the limit of small eccentricities, the inclination and node precessions are of order $\mathcal{O}(e)$, while the precessions of the perihelion and the mean anomaly are not defined when $e\rightarrow 0$ since they are of order $\mathcal{O}(e^{-1})$; on the other hand, the precession of the mean longitude $\lambda\doteq \varpi + \mathcal{M}$ is of order $\mathcal{O}(e)$. For other computation of the orbital effects, see \cite{Bail06}.
The generality of \rfr{dadt}-\rfr{dMdt} allows one to use them for sensitivity analyses in a variety of specific astronomical and astrophysical scenarios, where different $\{x,y\}$ reference planes and  orbital configurations of test particles appear.
\section{Results and conclusions}
By using \rfr{dodt} for Mercury, Venus and the Earth and the figures  of Table \ref{tavolafie} for their supplementary advances of perihelia, we solve for $s_x,s_y,s_z$ and obtain
\begin{align}
s_x \lb{dsx} & \doteq -\overline{s}^{01} =  (0.9\pm 1.5)\times 10^{-8}, \\ \nonumber \\
s_y \lb{dsy} & \doteq -\overline{s}^{02} = (-4 \pm 6)\times 10^{-9}, \\ \nonumber \\
s_z \lb{dsz} & \doteq -\overline{s}^{03} = (0.3\pm 1)\times 10^{-9}.
\end{align}
The forthcoming analysis of more data from the ongoing MErcury Surface, Space ENvironment,
GEochemistry, and Ranging (MESSENGER) mission \cite{messenger} to Mercury should allow one to improve such bounds in a near future. In view of that fact that. according to \rfr{dadt}-\rfr{dMdt}, the strongest signals occur for the closest particles to the central body, it would be desirable that supplementary advances of all the Keplerian orbital elements of Mercury will be produced in future ephemerides.

We stress that \rfr{dsx}-\rfr{dsz} should be regarded as a sort of predicted parameter sensitivity since, actually, Fienga et al. \cite{fienga011} did not model any Lorentz-violating terms in the INPOP010a ephemerides. Actually, in more refined analyses, it would be required, in principle, to explicitly include \rfr{accel} in the dynamical force models usually fitted to planetary observations, and solve for the $\overline{s}^{0j},\ j=1,2,3$ parameters in a dedicated global solution obtained by reprocessing the entire data set with such modified softwares. The use of a similar approach, including also LLR data, was envisaged in \cite{Batt07}; see also the considerations by Nordtvedt \cite{Nord} in a different context.

The Moon yields us a benchmark for testing the degree of reliability of our approach. Indeed, from the small eccentricity limit of \rfr{dodt}, it can  naively be posed
\eqi \dot\varpi \sim \rp{GMs}{c a^2 e}, \eqf from which one can infer
\eqi s\lesssim \rp{c a^2 e}{GM}\delta\dot\varpi.\lb{approx}\eqf Since for the lunar perigee it is \cite{Dick,Muller1,Williams,Muller2,Muller3}
\eqi \delta\dot\varpi \sim 0.1 \ {\rm mas\ yr^{-1}},\eqf
\rfr{approx} yields just
\eqi s \sim 10^{-7},\eqf in substantial agreement with the constraints of \rfr{llr1}-\rfr{llr2} obtained by Battat et al. \cite{Batt07} as the outcome of a fit of  modified models to LLR data.

Finally, it may be interesting to look at the Earth and the LAGEOS satellites. Indeed, by looking at, say, the nodes of both LAGEOS and LAGEOS II, and the perigee of LAGEOS II it is possible to constrain $\bds s$ as previously done with the perihelia of three planets. Generally speaking, a major source of systematic uncertainty in the knowledge of the orbit of a terrestrial spacecraft is represented by the mismodeling in the first even zonal coefficient $\overline{C}_{2,0}$ accounting for the terrestrial quadrupole mass moment because of the resulting relatively huge secular precessions of the node and the perigee \cite{Capde}. By assuming \cite{IERS} $\delta \overline{C}_{2,0}\sim 10^{-10}$, corresponding to uncertainties in the  orbital elements of LAGEOS and LAGEOS II of the order of $\sim 100-200$ mas yr$^{-1}$, it turns out
\begin{align}
s_x \lb{Tdsx} & \doteq -\overline{s}^{01} \lesssim  1.4\times 10^{-3}, \\ \nonumber \\
s_y \lb{Tdsy} & \doteq -\overline{s}^{02} \lesssim 2\times 10^{-4}, \\ \nonumber \\
s_z \lb{Tdsz} & \doteq -\overline{s}^{03} \lesssim 1.3 \times 10^{-3}.
\end{align}
It can be noticed that \rfr{Tdsx}-\rfr{Tdsz} are neither competitive with the planetary bounds of \rfr{dsx}-\rfr{dsz} nor with the lunar ones of \rfr{llr1}-\rfr{llr2}.
\bibliography{SMEbib,Anellobib,Operabib}{}

\end{document}